\documentclass[reprint,aps,prl,twocolumn]{revtex4-2}
\usepackage{graphicx}
\usepackage{xcolor}
\usepackage{mathtools}
\usepackage{comment}
\usepackage{braket}
\usepackage{ulem}
\usepackage{amsfonts}

\begin{document}

\title{Upper Bound on Locally Extractable Energy from Entangled Pure State\\ under Feedback Control}
\author{Kanji Itoh$^{1}$, Yusuke Masaki$^{1}$, and Hiroaki Matsueda$^{1,2}$\thanks{hiroaki.matsueda.c8@tohoku.ac.jp}}
\affiliation{
$^{1}$Department of Applied Physics, Graduate School of Engineering, Tohoku University, Sendai 980-8579, Japan\\
$^{2}$Center for Science and Innovation in Spintronics, Tohoku University, Sendai 980-8577, Japan
}
\date{\today}

\begin{abstract}
We introduce an effective thermodynamics for multipartite entangled pure states and derive an upper bound on extractable energy with feedback control from a subsystem under a local Hamiltonian. The inequality that gives the upper bound corresponds to the second law of information thermodynamics in our effective thermodynamics. In addition, we derive a more general bound that is determined only by an initial state and the local Hamiltonian. This bound gives an explicit relationship between the extractable energy and the entanglement structure of the initial state. We also investigate the tightness of the upper bounds and show that the bounds can be achieved in a simple example.
\end{abstract}

\maketitle

\textit{Introduction.}---In light of the rapid progress in quantum information technologies, there is an urgent need to elucidate the energy efficiency of quantum information processing~\cite{PRXQuantum.3.020101}. This challenge is important not only for applications in quantum technologies, but also for fundamental physics, as it requires deep understanding of the relationship between energy and quantum information.

The relationship between energy and information is one of the key issues in modern thermodynamics. In 2008, Sagawa and Ueda derived the second law of information thermodynamics~\cite{sagawa-ueda}. This law extends the applicability of the second law of thermodynamics to processes that involve measurement and feedback control. The extended law shows that feedback control can improve work extraction by exploiting information gain from the measurement. In the case of an isothermal process, the second law of information thermodynamics is expressed as
\begin{align}
    W_\mathrm{ext}\leq -\Delta F +\frac{1}{\beta} I_\mathrm{QC},\label{eq:2ndLaw}
\end{align}
where $W_\mathrm{ext}$ is the extracted work, $\Delta F$ is the change in the Helmholtz free energy of the system, $\beta$ is the inverse temperature and  $I_\mathrm{QC}$ is the QC-mutual information (or Groenewold-Ozawa information) defined as the information gain by measuring the system~\cite{Groenwold,Ozawa}. 
Although the inequality~(\ref{eq:2ndLaw}) was derived for quantum systems, the setup does not explicitly include entanglement, which is a necessary resource for quantum information processing. Around the last decade, as extensions of Eq.~\eqref{eq:2ndLaw}, several information thermodynamic inequalities have been derived in  entangled systems~\cite{tajima,PhysRevA.88.052319,PhysRevLett.121.120602,PhysRevLett.111.230402}.
These inequalities give quantitative relations between energy and quantum information in entangled systems. In the setup of the above studies, work is extracted from a system at the temperature of the heat bath.

On the other hand, \textit{local} energy extraction from multipartite quantum systems without thermal fluctuations has also been studied in various setups~\cite{PhysRevD.78.045006,hotta2010energy,PhysRevA.82.042329,PhysRevA.87.032313,trevison-hotta,PhysRevLett.123.190601,PhysRevB.107.075116,PhysRevLett.130.110801,PhysRevApplied.20.024051}. For such setups, general and quantitative energy-information relations, such as the second law of information thermodynamics, have not been established. Establishing such a relationship leads to a new theory of energy efficiency in various entanglement-based quantum protocols at low temperatures.

In this Letter, we show that locally extractable energy from a multipartite quantum system is bounded with quantum information derived from entanglement. In our setup, the whole system is in an entangled pure state as an initial state, and energy is extracted by measurement and feedback control performed on some parts of the system. In entangled quantum systems, even if a whole system is not thermally fluctuating, i.e., the whole system is in a pure state, a subsystem can be in a mixed state. One can consider that the mixed state is realized under an effective temperature characterized by the entanglement. Building on this idea, we divide the whole system into several local systems, and introduce an effective thermodynamics. The main result of this Letter is the derivation of two upper bounds on the extracted energy. One bound can be seen as the second law of information thermodynamics in our effective thermodynamics. The other bound is looser, but provides an explicit relation between energy and entanglement structure of the initial state of the energy extraction process. We also examine the tightness of the bounds with a simple four-qubit example.

\textit{Setup.}---Figure 1 schematically shows our setup. We divide the whole system into a system $S$, an ancilla $A$, and an environment $E$. The energy extraction protocol consists of measurement and feedback control. First, a projective measurement $P_{A}(\mu)$ is performed on the ancilla $A$, and the measurement result $\mu$ is obtained. Next, a feedback unitary operation $U_S(\mu)$ is performed on the system $S$, and the amount of energy $E_\mathrm{ext}$ is extracted from $S$. In the protocol, no operation is performed on the environment $E$. However, the environment $E$ is necessary to define temperature of our effective thermodynamics.

At the beginning of the energy extraction process, the state of the whole system $\rho_{SAE}^\mathrm{i}$ is an entangled pure state, i.e.,
$\rho_{SAE}^\mathrm{i}\equiv \ket{\psi}_{SAE}\bra{\psi}_{SAE}$, where $\ket{\psi}_{SAE}$ is an entangled state vector.
In this case, the system $S$ is in a mixed state due to the entanglement. The von Neumann entropy of this mixed state represents information derived only from entanglement and is called entanglement entropy. The definition of the entanglement entropy of the system $S$ is given by $S(\rho_S^\mathrm{i})\equiv -\mathrm{tr}\rho_S^\mathrm{i} \log \rho_S^\mathrm{i}$,
where $\rho_S^\mathrm{i} \equiv \mathrm{tr}_{AE} \rho_{SAE}^\mathrm{i}$ is the reduced density matrix of $\ket{\psi}_{SAE}$ on the system $S$.
Here and hereafter, $S(\rho)$ denotes the von Neumann entropy of a density matrix $\rho$.

\begin{figure}[t]
\begin{center}
\includegraphics[width=8cm]{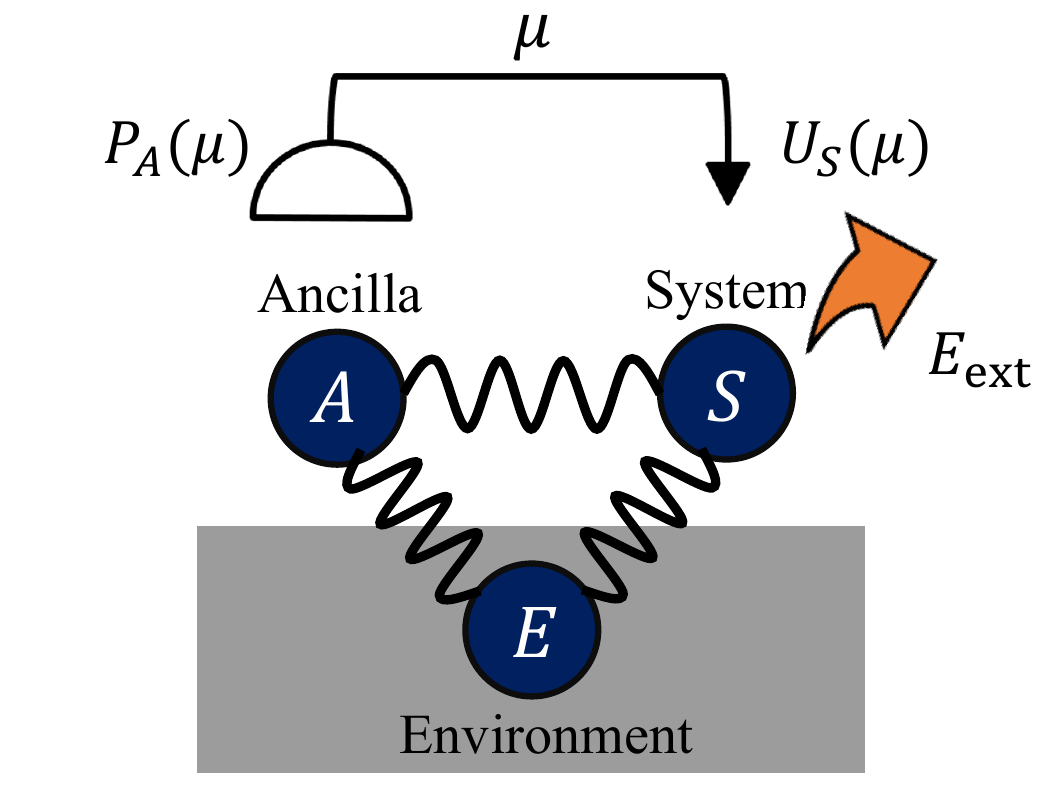}
\end{center}
\vspace{-0.3cm}
\caption{Illustration of the setup. The whole system is initially in a entangled pure state, and is divided into the system $S$, ancilla $A$, and environment $E$. As the energy extraction protocol, projective measurement $P_{A}(\mu)$ is performed on $A$, and feedback unitary operation $U_S(\mu)$ is performed on $S$.}
\label{fig:setup}
\end{figure}

After the measurement with a result $\mu$, the state of the whole system is written as $\rho_{SAE}^\mathrm{m}(\mu)=\frac{1}{p_\mu}P_{A}(\mu)\rho_{SAE}^\mathrm{i}P_{A}(\mu)$,
where $p_\mu\equiv \bra{\psi}P_{A}(\mu)\ket{\psi}_{SAE}$ is the probability of obtaining a result $\mu$. The reduced density matrix of the system $S$ after obtaining the result $\mu$ is expressed by $\rho_S^\mathrm{m}(\mu)=\mathrm{tr}_{AE} \rho_{SAE}^\mathrm{m}(\mu)$.
In our setup, the information gain of the system $S$ due to the measurement is calculated as the average reduction of the entanglement entropy: $I_\mathrm{QC}\equiv S(\rho_S^\mathrm{i})-\sum_\mu p_\mu S\left(\rho_S^\mathrm{m}(\mu)\right)$.

Next, the feedback unitary operation $U_S(\mu)$ is performed on the system $S$. The final state of the whole system is represented by $\rho_{SAE}^\mathrm{f}=\sum_\mu U_S(\mu) P_{A}(\mu) \rho_{SAE}^\mathrm{i} P_{A}(\mu) U_S^\dagger(\mu)$. 
Thus, the final state of the system $S$ is given by $\rho_S^\mathrm{f}=\mathrm{tr}_{AE}\rho_{SAE}^\mathrm{f}=\sum_\mu p_\mu U_S(\mu)\rho_S^\mathrm{m}(\mu)U_S^\dagger(\mu)$.
The extracted energy is given by the energy reduction of the system $S$ during the protocol:
\begin{align}
    E_\mathrm{ext}\equiv E_S^\mathrm{i}-E_S^\mathrm{f},\label{eq:def_Eext}
\end{align}
where $E_S^\mathrm{i}\equiv \mathrm{tr}\rho_S^\mathrm{i}H_S$ and $E_S^\mathrm{f}\equiv\mathrm{tr}\rho_S^\mathrm{f}H_S$ are the energy expectation values of the initial and final states of $S$, respectively, and $H_S$ is the local Hamiltonian of $S$ at the beginning and the end of the protocol.
Note that the \textit{maximally} extracted energy by feedback unitary operations under a fixed projective measurement is called daemonic ergotropy~\cite{daemonic_ergotropy1}. In this Letter, we will derive \textit{upper bounds} on the extracted energy. As will be mentioned later, when our bound can be achieved, the bound is equal to the daemonic ergotropy; in this case, our bound gives information thermodynamic expression of the daemonic ergotropy for pure states.

\textit{Effective Thermodynamics and Upper Bounds.}---Let us introduce an  effective thermodynamics for entangled pure states. In our effective thermodynamics, the energy extraction protocol is considered as a nonequilibrium process of the system $S$ at an effective temperature, which is defined by the entanglement between the system $S$ and the environment $E$. In the following, we derive a second law-like inequality of the effective thermodynamics.

First, we define the effective thermal equilibrium state. As mentioned above, the system $S$ is in a mixed state due to the entanglement. Thus, we consider the system $S$ to be at an effective temperature $1/\beta_\mathrm{eff}>0$, for which we define the thermal equilibrium state as
\begin{align}
    \sigma_S\equiv \frac{1}{Z}e^{-\beta_\mathrm{eff}H_S},
\end{align}
where $Z\equiv \mathrm{tr}~e^{-\beta_\mathrm{eff}H_S}$ is the partition function. The thermal equilibrium state $\sigma_S$ depends only on $\beta_\mathrm{eff}$, which is defined by the entanglement between the system $S$ and the environment $E$. We postpone the explicit definition of $\beta_\mathrm{eff}$ until Eq.~(\ref{eq:condition}).

Next, we rewrite the extracted energy $E_\mathrm{ext}$ with quantities based on the effective thermodynamics. In the effective thermodynamics, the initial state $\rho_S^\mathrm{i}$ and the final state $\rho_S^\mathrm{f}$ of the protocol are considered as nonequilibrium states at the effective temperature $\beta_\mathrm{eff}$. Thus, we define nonequilibrium free energy~\cite{Esposito_2011} of the states $\rho_S^\alpha$ for $\alpha=\mathrm{i,f}$ as
\begin{align}
    \mathcal{F}(\rho_S^\alpha;H_S)&\equiv E_S^\alpha-\frac{1}{\beta_\mathrm{eff}}S(\rho_S^\alpha)\nonumber\\
    &=F(\sigma_S)+\frac{1}{\beta_\mathrm{eff}}D(\rho_S^\alpha||\sigma_S),\label{eq:def_noneqF}
\end{align}
where $F(\sigma_S)\equiv -\frac{1}{\beta_\mathrm{eff}}\log Z$ is the Helmholtz free energy of $\sigma_S$ and $D(\rho_S^\alpha||\sigma_S)$ is the Kullback-Leibler (KL) divergence (or quantum relative entropy).
Using the nonequilibrium free energy, $E_\mathrm{ext}$ defined by Eq.~(\ref{eq:def_Eext}) can be rewritten as
\begin{align}
    E_\mathrm{ext}=-\Delta \mathcal{F}_S - \frac{1}{\beta_\mathrm{eff}} \Delta S_S,\label{eq:Eext2}
\end{align}
where $\Delta \mathcal{F}_S\equiv \mathcal{F}(\rho_S^\mathrm{f};H_S)-\mathcal{F}(\rho_S^\mathrm{i};H_S)$ is the change in the nonequilibrium free energy and $\Delta S_S\equiv S(\rho_S^\mathrm{f})-S(\rho_S^\mathrm{i})$ is the change in the von Neumann entropy. Note that $\Delta S_S$ is always zero when $U_S(\mu)$ is not feedback control, i.e., $U_S(\mu)$ is independent of $ \mu$.

Let us now move on to the derivation of upper bounds on $E_\mathrm{ext}$. First, from the positivity of Kullback-Leibler divergence $D(\rho_S^\mathrm{f}||\sigma_S)$, we obtain
\begin{align}
    -\Delta \mathcal{F}_S\leq\mathcal{F}(\rho_S^\mathrm{i};H_S)-F(\sigma_S).\label{eq:ineq_F}
\end{align}
The equality holds if and only if $\rho_S^\mathrm{f}=\sigma_S$.
Next, we focus on the entropy term in Eq.~(\ref{eq:Eext2}). Using the convexity of entropy, we have
\begin{align}
    S(\rho_S^\mathrm{f})\geq\sum_\mu p_\mu S(\rho_S^\mathrm{m}(\mu))
    \geq E_\mathrm{F}^{S\textit{-}E}(\rho_{SE}^\mathrm{i}),\label{eq:ineq_S}
\end{align}
where $\rho_{SE}^\mathrm{i}\equiv \mathrm{tr}_A \rho_{SAE}^\mathrm{i}$ and $E_\mathrm{F}^{S\textit{-}E}(\rho_{SE}^\mathrm{i})$ is entanglement of formation~\cite{EOF}, representing mixed state entanglement between the system $S$ and the environment $E$. 
In our setup, $E_\mathrm{F}^{S\textit{-}E}(\rho_{SE}^\mathrm{i})$ can be expressed as
\begin{align}
    E_\mathrm{F}^{S\textit{-}E}(\rho_{SE}^\mathrm{i}) \equiv \min_{\{P_A(\mu)\}} \sum_\mu p_\mu S(\rho_S^\mathrm{m}(\mu)),
\end{align}
where the minimization is performed over all sets of possible orthogonal projection operators. This minimization is equivalent to optimizing the representation of the mixed state $\rho_{SE}^\mathrm{i}$ by mixture of pure states: $\sum_\mu p_\mu \ket{\phi_\mu}_{SE}\bra{\phi_\mu}_{SE} = \rho_{SE}^\mathrm{i}$, where we put $\ket{\psi}_{SAE}=\sum_\mu \sqrt{p_\mu} \ket{a_\mu}_A \otimes \ket{\phi_\mu}_{SE}$ and $P_A(\mu)=\ket{a_\mu}_A \bra{a_\mu}_A$. Note that $\{\ket{\phi_\mu}\}$ need not be an orthogonal basis set. 
From Eq.~(\ref{eq:ineq_S}), we obtain
\begin{align}
    -\Delta S_S \leq I_\mathrm{QC} \leq \overleftarrow{\mathcal{E}}_{SA},\label{eq:ineq_S2}
\end{align}
where we define the new quantity $\overleftarrow{\mathcal{E}}_{SA}$ which represents asymmetric entanglement between $S$ and $A$. The quantity $\overleftarrow{\mathcal{E}}_{SA}$ is defined as entanglement between $S$ and $AE$ minus entanglement between $S$ and $E$:
\begin{align}
    \overleftarrow{\mathcal{E}}_{SA}\equiv S(\rho_S^\mathrm{i})-E_\mathrm{F}^{S\textit{-}E}(\rho_{SE}^\mathrm{i}).\label{eq:def_Esa}
\end{align}
Plugging inequalities~(\ref{eq:ineq_F}) and (\ref{eq:ineq_S2}) in Eq.~(\ref{eq:Eext2}) yields upper bounds on $E_\mathrm{ext}$:
\begin{align}
    E_\mathrm{ext} &\leq \mathcal{F}(\rho_S^\mathrm{i};H_S)-F(\sigma_S)+\frac{1}{\beta_\mathrm{eff}}I_\mathrm{QC}\nonumber\\
    &\leq \mathcal{F}(\rho_S^\mathrm{i};H_S)-F(\sigma_S)+\frac{1}{\beta_\mathrm{eff}} \overleftarrow{\mathcal{E}}_{SA}.
    \label{eq:bound}
\end{align} 
The first bound in Eq.~(\ref{eq:bound}) depends on the choice of the set of projection operators $\left\{P_A(\mu)\right\}$ as well as $I_\mathrm{QC}$. When this bound can be achieved, the bound is the daemonic ergotropy itself. The second bound in Eq.~(\ref{eq:bound}) is determined only by the initial state $\ket{\psi}_{SAE}$ and the local Hamiltonian $H_S$.

A necessary condition for the both of the equalities in Eq.~(\ref{eq:bound}) to hold is given by
\begin{align}
    S(\sigma_S)=E_\mathrm{F}^{S\textit{-}E}(\rho_{SE}^\mathrm{i}).\label{eq:condition}
\end{align}
To make the bounds~(\ref{eq:bound}) achievable, we define the effective temperature $1/\beta_\mathrm{eff}$ so that Eq.~(\ref{eq:condition}) holds. By this definition, $1/\beta_\mathrm{eff}$ is always uniquely determined for $H_S\neq0$. This can be confirmed as follows. The left side of Eq.~(\ref{eq:condition}) decreases monotonically with respect to $\beta_\mathrm{eff}$ from $S(\sigma_S)=\log(\dim\sigma_S)$ for $\beta_\mathrm{eff}\to0$ to $S(\sigma_S)\to0$ for $\beta_\mathrm{eff}\to \infty$. On the other hand, $E_\mathrm{F}^{S\textrm{-}E}(\rho_{SE}^\mathrm{i})$ satisfies $0\leq E_\mathrm{F}^{S\textrm{-}E}(\rho_{SE}^\mathrm{i}) \leq \min{\left\{\log{(\dim\rho_S^\mathrm{i})}, \log{(\dim\rho_E^\mathrm{i})}\right\}}\leq \log{(\dim{\sigma_S})}$, where $\rho_E^\mathrm{i}\equiv \mathrm{tr}_{SA}\rho_{SAE}^\mathrm{i}$.
Therefore, there exists only one value of $\beta_\mathrm{eff}$ satisfying Eq.~(\ref{eq:condition}).
Under the above definition of $\beta_\mathrm{eff}$, $E_\mathrm{ext}$ reaches the second bound in Eq.~(\ref{eq:bound}) only if the eigenvalues of $\rho_S^\mathrm{m}(\mu)$ that maximize $I_\mathrm{QC}$ are independent of the measurement results $\mu$. The proof of this is given in Supplemental Material. We will show that $E_\mathrm{ext}$ can actually reach the bound in a simple model when the above condition is satisfied.

Let us discuss the physical meaning of the bounds in Eq.~(\ref{eq:bound}). Using Eq.~(\ref{eq:def_noneqF}), the bounds can be rewritten more simply as
\begin{align}
E_\mathrm{ext}&\leq\frac{D(\rho_S^\mathrm{i}||\sigma_S)+I_\mathrm{QC}}{\beta_\mathrm{eff}}\leq\frac{D(\rho_S^\mathrm{i}||\sigma_S)+\overleftarrow{\mathcal{E}}_{SA}}{\beta_\mathrm{eff}}.\label{eq:bound_re}
\end{align}
The KL divergence $D(\rho_S^\mathrm{i}||\sigma_S)$, which represents how different the initial state of the system $S$ is from the equilibrium state $\sigma_S$, is a resource of the energy extraction. Besides, the information obtained from the measurement can improve the extracted energy. This situation is similar to the case in the previous studies of nonequilibrium information thermodynamic process~\cite{Esposito_2011,parrondo-horowitz-sagawa}. Thus, the first bound in Eq.~(\ref{eq:bound_re}) (or Eq.~(\ref{eq:bound})) can be considered as the second law of information thermodynamics in our effective thermodynamics.

Obviously from Eq.~(\ref{eq:def_Esa}), the whole entanglement of the system $S$ in the initial state can be decomposed into the $S$-$A$ entanglement $\overleftarrow{\mathcal{E}}_{SA}$ and the $S$-$E$ entanglement $E_\mathrm{F}^{S\textit{-}E}(\rho_{SE}^\mathrm{i})$. The former entanglement $\overleftarrow{\mathcal{E}}_{SA}$ can be considered as the informative resource for the energy extraction, because $\overleftarrow{\mathcal{E}}_{SA}$ is the upper bound on the information gain $I_\mathrm{QC}$ and $I_\mathrm{QC}$ can contribute positively to the energy extraction. On the other hand, the latter entanglement $E_\mathrm{F}^{S\textit{-}E}(\rho_{SE}^\mathrm{i})$ is the origin of the temperature in our effective thermodynamics. Therefore, the second bound in Eq.~(\ref{eq:bound_re}) gives the explicit relationship between the extracted energy $E_\mathrm{ext}$ and the entanglement structure of the initial state $\ket{\psi}_{SAE}$.

\textit{Example.}---Let us examine the tightness of the bound~(\ref{eq:bound_re}) with a simple example of a four-qubit system consisting of qubits 1, 2, 3, and 4. Energy is extracted from qubit 1 by a projective measurement performed on qubit 2 and a feedback unitary operation performed on qubit 1. Thus, qubits 1 and 2 correspond to the system $S$ and the ancilla $A$, respectively, and qubits 3 and 4 correspond to the environment $E$. 
The local Hamiltonian of qubit 1 is defined by $H_S\equiv \sigma_1^z$, where $\sigma_1^z$ is the $z$ component of the Pauli matrices for qubit 1. We define the initial state $\ket{\psi}$ with a parameter $\eta$ as follows:
\begin{align}
    \ket{\psi} &\equiv \sqrt{\frac{1+\eta}{14}}(2\ket{\uparrow\uparrow\uparrow\uparrow}+\ket{\uparrow\uparrow\uparrow\downarrow}+\ket{\uparrow\downarrow\uparrow\uparrow}+\ket{\uparrow\downarrow\uparrow\downarrow}) \nonumber\\ +&\sqrt{\frac{1-\eta}{14}}(\ket{\downarrow\uparrow\downarrow\uparrow}+\ket{\downarrow\uparrow\downarrow\downarrow}-\ket{\downarrow\downarrow\downarrow\uparrow}-2\ket{\downarrow\downarrow\downarrow\downarrow}),
    \label{eq:example}
\end{align}
where we use the abbreviated notation such as $\ket{\uparrow\uparrow\uparrow\uparrow}=\ket{\uparrow}_1 \otimes \ket{\uparrow}_2 \otimes \ket{\uparrow}_3 \otimes \ket{\uparrow}_4$. 
The spin bases of qubit $i$ $(i=1,\cdots,4)$, $\ket{\uparrow}_{i}$ and $\ket{\downarrow}_{i}$, are the eigenstates of $\sigma_i^z$ as $\sigma_i^z \ket{\uparrow}_{i} = \ket{\uparrow}_{i}$ and $\sigma_i^z \ket{\downarrow}_{i} = -\ket{\downarrow}_{i}$.
Since the entanglement of the qubit 1 vanishes for $\eta=\pm 1$, we consider $-1<\eta<1$. When $\eta=0$, the eigenvalues of $\rho_S^\mathrm{m}(\mu)$ are independent of $\mu$, i.e., the necessary condition for $E_\mathrm{ext}$ to reach the upper bound is satisfied. At this value of $\eta$, the reduced density matrix of Eq.~(\ref{eq:example}) for qubit 1 is the  highly symmetric state, which is invariant to any unitary operation: $\rho_S^\mathrm{i}=(\ket{\uparrow}_1\bra{\uparrow}_1+\ket{\downarrow}_1\bra{\downarrow}_1)/2$. This symmetry is broken for $|\eta|>0$.

\begin{figure}[t]
\begin{center}
\includegraphics[width=9cm]{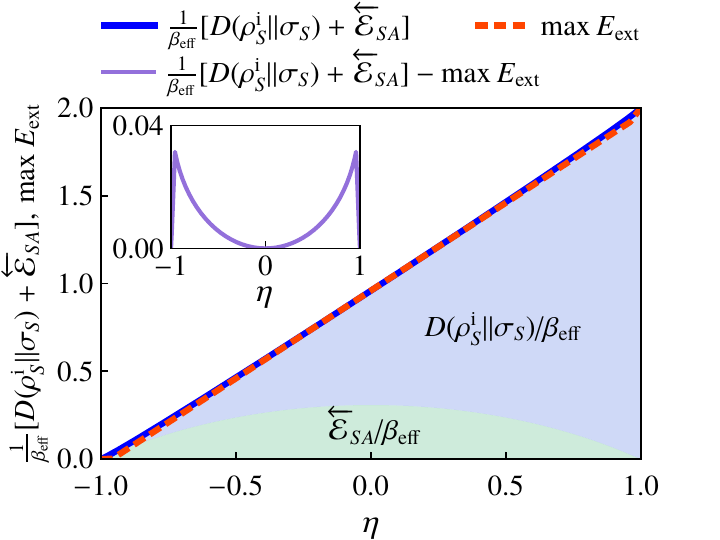}
\end{center}
\vspace{-0.3cm}
\caption{Tightness of the second bound in Eq.~(\ref{eq:bound_re}). The bound is compared to the maximally extracted energy, and is displayed separately for the contributions of the free energy term $D(\rho_S^\mathrm{i}||\sigma_S)/\beta_\mathrm{eff}$ (the blue area) and entanglement term $\overleftarrow{\mathcal{E}}_{SA}/\beta_\mathrm{eff}$ (the green area). The inset shows the difference between the upper bound and maximally extracted energy}.
\label{fig:tightness}
\end{figure}

Figure~\ref{fig:tightness} shows the tightness of the second bound in Eq.~(\ref{eq:bound_re}) in the above model. Both the upper bound and the maximally extracted energy increase monotonically with respect to $\eta$. This is because the energy of the initial state of qubit 1 is higher for larger $\eta$. This increase of the initial state energy corresponds to the increase of the nonequilibrium free energy of the initial state in terms of the effective thermodynamics. In contrast to the $\eta$-dependence of the free energy term $D(\rho_S^\mathrm{i}||\sigma_S)/\beta_\mathrm{eff}$, the entanglement term $\overleftarrow{\mathcal{E}}_{SA}/\beta_\mathrm{eff}$ decreases monotonically with $|\eta|$. As shown in the inset, the difference between the upper bound and the maximally extracted energy is always less than $0.04$ even though the magnitude of the bound and the maximally extracted energy increase up to $2$. Therefore, the bound is tight except for $\eta \approx -1$. In addition, the maximally extracted energy reaches the bound at $\eta=0$. Thus, our bound is achievable for highly symmetric initial state in this entangled four qubit model. The details of the calculation are given in Supplemental Material.

\textit{Conclusion.}---We introduced effective thermodynamics for entangled multipartite pure states and derived two upper bounds on locally extractable energy with feedback control. One bound corresponds to the second law of information thermodynamics in our effective thermodynamics, and the other bound shows the explicit relationship between energy and the entanglement structure of the initial state. We also investigated the tightness of the bounds with a simple model, and showed that the bounds are achievable for highly symmetric initial states.

Our results go beyond the previous framework of information thermodynamics, the validity range of which has recently been verified~\cite{minagawa2023universal}. The results open the door to entanglement-based information thermodynamics without thermal fluctuations and lead to a fundamental theory for the energy efficiency of quantum information processing at low temperatures.

This work was supported by JST SPRING, Grant Number JPMJSP2114, JSPS KAKENHI Grant Numbers JP24K06878, JP24K00563, JP24K02948, JP24K17000, JP23K22492, JP21H04446, JP21K03380, and GP-Spin and CSIS in Tohoku University.

\bibliography{bibl.bib}

\begin{widetext}

\vspace{1.5cm}

\setcounter{equation}{0}
\newtagform{Supplement}[]{(S.}{)}
\usetagform{Supplement}

\begin{center}
 \textbf{{\large Supplemental Material}}
\end{center}

\vspace{3mm}

In this Supplemental Material, we present a proof of the equality condition of the upper bound. In addition, we provide details on the calculation of the extractable energy and the upper bound in the simple example presented in the main text.

\section{Proof of the Equality Condition}
As mentioned before Eq.~(13) in the main text, when the effective temperature $1/\beta_\mathrm{eff}$ is defined so that Eq.~(12) holds, a necessary condition for $E_\mathrm{ext}$ to be reach the second bound in Eq.~(11) (or Eq.~(13)) in the main text is given by the following theorem.

\textbf{Theorem.} \textit{Let $\rho_S^\mathrm{m}(\mu)$ be post measurement states that maximizes $I_\mathrm{QC}$. Then $\max E_\mathrm{ext}=\frac{1}{\beta_\mathrm{eff}}[D(\rho_S^\mathrm{i}||\sigma_S)+\overleftarrow{\mathcal{E}}_{SA}]$ holds only if the eigenvalues of $\rho_S^\mathrm{m}(\mu)$ are independent of $\mu$.}

\textbf{Proof.}
Suppose $\max E_\mathrm{ext}=\frac{1}{\beta_\mathrm{eff}}[D(\rho_S^\mathrm{i}||\sigma_S)+\overleftarrow{\mathcal{E}}_{SA}]$. Then the equalities in Eq.~(7) in the main text hold, i.e., there exists $U_S(\mu)$ which satisfies the equality of $
S(\rho_S^\mathrm{f}) \ge \sum_\mu p_\mu S(\rho_S^\mathrm{m}(\mu))$ for $\rho_S^\mathrm{m}(\mu)$ that maximizes $I_\mathrm{QC}$. This inequality is derived by using positivity of KL divergence:
\begin{align}
    S(\rho_S^\mathrm{f}) &= -\mathrm{tr} \rho_S^\mathrm{f} \log \rho_S^\mathrm{f} \nonumber\\
    &= -\sum_\mu  p_\mu \mathrm{tr}  U_S(\mu)\rho_S^\mathrm{m}(\mu)U_S^\dagger(\mu) \log \rho_S^\mathrm{f} \nonumber\\
    &\ge -\sum_\mu p_\mu \mathrm{tr}  U_S(\mu)\rho_S^\mathrm{m}(\mu)U_S^\dagger(\mu) \log U_S(\mu)\rho_S^\mathrm{m}(\mu)U_S^\dagger(\mu) \nonumber\\
    &= \sum_\mu p_\mu S(\rho_S^\mathrm{m}(\mu)),
\end{align}
where $\rho_S^{\mathrm{f}} = \sum_{\mu}p_\mu \mathrm{tr} U_S(\mu)\rho_{S}^{\mathrm{m}}(\mu)U_{S}^\dagger(\mu)$ is used to obtain the second line. In the third line,
the equality holds if and only if $U_S(\mu) \rho_S^\mathrm{m}(\mu)U_S^\dagger(\mu)=\rho_S^\mathrm{f}$ for all $\mu$. Since $U_S(\mu)$ does not change eigenvalues of $\rho_S^\mathrm{m}(\mu)$, the equality condition implies that the eigenvalues of $\rho_S^\mathrm{m}(\mu)$ are independent of $\mu$. $\square$

Note that the converse of the above theorem is not necessarily true. If the eigenvalues of $\rho_S^\mathrm{m}(\mu)$ are independent of $\mu$, there exist $P_A(\mu)$ and $U_S(\mu)$ which satisfy
\begin{align}
    S(\rho_S^\mathrm{f})=\sum_\mu p_\mu S(\rho_S^\mathrm{m}(\mu))=S(\rho_S^\mathrm{m}(\mu))=E_\mathrm{F}^{S\textit{-}E}(\rho_{SE}^\mathrm{i})=S(\sigma_S).\label{eq:converse}
\end{align}
However, since $S(\rho_S^\mathrm{m}(\mu))$=$S(\sigma_S)$ does not imply that the eigenvalues of $\rho_S^\mathrm{m}(\mu)$ and $\sigma_S$ are equal, Eq.~(S.\ref{eq:converse}) does not imply $\rho_S^\mathrm{f}=\sigma_S$, which is the necessary condition for $E_\mathrm{ext}$ to reach the upper bound. It is an open question whether there exist initial states $\ket{\psi}_{SAE}$ that cannot reach the upper bound and whose post-measurement states $\rho_S^\mathrm{m}(\mu)$'s have eigenvalues that are independent of the measurement results.

\section{Example}
In this section, we investigate in what initial states the above equality condition of the bound is satisfied in a four qubit example. We also investigate the tightness of the bound and provide details of the calculation. As a simple example, we consider a multipartite system consists of qubits 1, 2, 3, and 4. Measurement is performed on qubit 2, and feedback control is performed on qubit 1. Thus, qubit 1 is the system $S$, qubit 2 is the ancilla $A$, and  qubits 3 and 4 are the environment $E$.
For simplicity, we define local Hamiltonian of qubit 1 as 
\begin{align}
    H_S\equiv \sigma_1^z
    =\begin{pmatrix}
        1 & 0  \\
        0 & -1 \\
    \end{pmatrix},
\end{align}
and the initial state as
\begin{align}
    \ket{\psi}_{SAE}\equiv &a\ket{\uparrow\uparrow\uparrow\uparrow}_{1234}+b\ket{\uparrow\uparrow\uparrow\downarrow}_{1234}+c\ket{\uparrow\downarrow\uparrow\uparrow}_{1234}+d\ket{\uparrow\downarrow\uparrow\downarrow}_{1234}\nonumber\\&+e\ket{\downarrow\uparrow\downarrow\uparrow}_{1234}+f\ket{\downarrow\uparrow\downarrow\downarrow}_{1234}+g\ket{\downarrow\downarrow\downarrow\uparrow}_{1234}+h\ket{\downarrow\downarrow\downarrow\downarrow}_{1234}.\label{eq:psi_example}
\end{align}
The initial state of qubit 1 is written as
\begin{align}
\rho_S^\mathrm{i}&=\mathrm{tr}_{AE}\ket{\psi}_{SAE}\bra{\psi}_{SAE}\nonumber\\
&=\begin{pmatrix}
    a^2+b^2+c^2+d^2 & 0 \\
    0 & e^2+f^2+g^2+h^2 
\end{pmatrix}.
\end{align}
We chose the initial state so that there is no need of diagonalization in the computation of $\max E_\mathrm{ext}$ and the upper bounds on $E_\mathrm{ext}$. In this model, projective measurement $P_A(\mu)$ and feedback unitary operation $U_S(\mu)$ is written as
\begin{align}
    P_A(\mu)&= \frac{1}{2}[1+(-1)^\mu \vec{n}\cdot\vec{\sigma}]\nonumber\\&=\frac{1}{2}[1+(-1)^\mu (n_x\sigma_x+n_y\sigma_y+n_z\sigma_z),\\
    U_S(\mu)&=\cos{\theta_\mu}+i \vec{u}_\mu\cdot\vec{\sigma}\sin{\theta_\mu}\nonumber\\&=\cos{\theta_\mu}+i (u_{x\mu}\sigma_x+u_{y\mu}\sigma_y+u_{z\mu}\sigma_z) \sin{\theta_\mu},
\end{align}
where $\mu=0,1$ is the measurement results, and $\vec{n}\equiv (n_x, n_y, n_z)$ and $\vec{u}_\mu\equiv (u_{x\mu}, u_{y\mu}, u_{z\mu})$ are unit vectors, i.e.,  $n_x^2+n_y^2+n_z^2=1$ and $u_{x\mu}^2+u_{y\mu}^2+u_{z\mu}^2=1$.

Next, we examine the equality condition stated in the theorem. The post measurement state $\rho_S^\mathrm{m}(\mu)$ is computed as
\begin{align}
    \rho_S^\mathrm{m}(\mu)=&\frac{1}{p_\mu}\mathrm{tr}_{AE}P_A(\mu)\ket{\psi}_{SAE}\bra{\psi}_{SAE}P_A(\mu)\nonumber\\
    =&\frac{1}{2p_\mu}\left[a^2+b^2+c^2+d^2+(-1)^\mu\left\{2(ac+bd)n_x+(a^2+b^2-c^2-d^2)n_z\right\}\right]\ket{\uparrow}_1 \bra{\uparrow}_1 \nonumber\\
    &+\frac{1}{2p_\mu}\left[e^2+f^2+g^2+h^2+(-1)^\mu\left\{2(eg+fh)n_x+(e^2+f^2-g^2-h^2)n_z\right\}\right]\ket{\downarrow}_1 \bra{\downarrow}_1,
\end{align}
where $p_\mu$ is the probability of obtaining a result $\mu$:
\begin{align}
    p_\mu&=\bra{\psi}P_A(\mu)\ket{\psi}\nonumber\\
    &=\frac{1}{2}+(-1)^\mu \left[(ac+bd+eg+fh)n_x+\frac{1}{2}(a^2+b^2-c^2-d^2+e^2+f^2-g^2-h^2)n_z\right].
\end{align}
Thus, if following equations hold, the eigenvalues of $\rho_S^\mathrm{m}(\mu)$'s are independent of $\mu$ for any measurement.
\begin{subequations}
\begin{align}
    a^2+b^2+c^2+d^2&=e^2+f^2+g^2+h^2=\frac{1}{2},\label{eq:condition_S_a}\\
    ac+bd&=-(eg+fh),\\
    a^2+b^2-c^2-d^2&=-(e^2+f^2-g^2-h^2).
\end{align}\label{eq:condition_S}%
\end{subequations}
If Eq.~(S.\ref{eq:condition_S_a}) holds, the initial state of qubit 1 is invariant to any unitary operation: $\rho_S^\mathrm{i}=(\ket{\uparrow}_1\bra{\uparrow}_1+\ket{\downarrow}_1\bra{\downarrow}_1)/2$. This suggests that the upper bound is tight for highly symmetric initial states.

In the following, we provide detailed calculations of the upper bound and the extractable energy for a specific choice of the coefficients with parameter $\eta$: $a/2=b=c=d=\sqrt{1+\eta}/\sqrt{14}$ and $e=f=-g=-h/2=\sqrt{1-\eta}/\sqrt{14}$. It is also shown analytically that the upper bound is achievable when Eq.~(S.\ref{eq:condition_S}) holds.
The initial state is written down as
\begin{align}
    \ket{\psi}_{SAE} =& \sqrt{\frac{1+\eta}{14}}(2\ket{\uparrow\uparrow\uparrow\uparrow}_{1234}+\ket{\uparrow\uparrow\uparrow\downarrow}_{1234}+\ket{\uparrow\downarrow\uparrow\uparrow}_{1234}+\ket{\uparrow\downarrow\uparrow\downarrow}_{1234}) \nonumber\\ &+\sqrt{\frac{1-\eta}{14}}(\ket{\downarrow\uparrow\downarrow\uparrow}_{1234}+\ket{\downarrow\uparrow\downarrow\downarrow}_{1234}-\ket{\downarrow\downarrow\downarrow\uparrow}_{1234}-2\ket{\downarrow\downarrow\downarrow\downarrow}_{}1234),
    \label{example}
\end{align}
where $-1<\eta<1$.
When $\eta=0$, the conditions of Eq.~(S.\ref{eq:condition_S}) hold. First, we calculate maximally extracted energy $\max E_\mathrm{ext}.$ At the beginning of the energy extraction process, the reduced density matrix of qubit 1 is
\begin{align}
    \rho_S^\mathrm{i}&=
    \mathrm{tr}_{AE} \ket{\psi}_{SAE}\bra{\psi}_{SAE}\nonumber\\
    &=\frac{1}{2}
    \begin{pmatrix}
        1+\eta & 0 \\
        0 & 1-\eta 
    \end{pmatrix},
\end{align}
and thus the energy expectation value in the initial state of qubit 1 is calculated as
\begin{align}
    E_S^\mathrm{i}\equiv\mathrm{tr}\rho_S^\mathrm{i}H_S=\eta.
\end{align}
The post measurement state of qubit 1 with a result $\mu$ is
\begin{align}
    \rho_S^\mathrm{m}(\mu)
    &=\frac{1}{2p_\mu}
    \begin{pmatrix}
        \frac{1+\eta}{14}[7+(-1)^\mu 3(2n_x+n_z)] & 0 \\
        0 & \frac{1-\eta}{14}[7+(-1)^{\mu+1} 3(2n_x+n_z)]
    \end{pmatrix}\nonumber\\
    &\eqcolon
    \begin{pmatrix}
        \lambda_\uparrow(\mu) & 0 \\
        0 & \lambda_\downarrow(\mu)
    \end{pmatrix},\\
    p_\mu&=\frac{1}{14}\left[7+(-1)^\mu 3\eta(2n_x+n_z)\right].
\end{align}
Thus, the maximally extracted energy can be calculated as
\begin{align}
    \max{E_\mathrm{ext}}
    &=E_S^\mathrm{i}-\min_{\{P_A(\mu)\}, \{U_S(\mu)\}}{E_S^\mathrm{f}}\nonumber\\
    &=\eta+\max_{\{P_A(\mu)\}, \{U_S(\mu)\}}{\left\{-\sum_\mu p_\mu \mathrm{tr}U_S(\mu) \rho_S^\mathrm{m}(\mu) U_S^\dagger(\mu) H_S\right\}}\nonumber\\
    &=\eta+\max_{\{P_A(\mu)\}}\sum_\mu p_\mu |\lambda_\uparrow(\mu)-\lambda_\downarrow(\mu)|\nonumber\\
    &=\eta+\frac{1}{14}\max_{\{P_A(\mu)\}}\left\{|7\eta+3(2n_x+n_z)|+|7\eta-3(2n_x+n_z)|\right\}\nonumber\\
    &=\begin{cases}
      0 & \text{for $-1 < \eta \leq -\frac{3\sqrt{5}}{7}$},\\
      \eta+\frac{3\sqrt{5}}{7} & \text{for $-\frac{3\sqrt{5}}{7} \leq \eta \leq \frac{3\sqrt{5}}{7}$},\\
      2\eta & \text{for $\frac{3\sqrt{5}}{7} \leq \eta <1$},
   \end{cases}
\end{align}
where we used $-\sqrt{5} \leq 2n_x+n_z \leq \sqrt{5}$. The extracted energy is maximized when
\begin{align}
    \begin{cases}
        U_S(0)=U_S(1)=I_S &\text{for } -1< \eta \leq - \frac{3\sqrt{5}}{7},\\
        \text{$2n_x+n_z=- \sqrt{5}$, $U_S(0)=I_S$, $U_S(1)=\sigma_1^x$, or }  \text{$2n_x+n_z= \sqrt{5}$, $U_S(0)=\sigma_1^x$, $U_S(1)=I_S$} &\text{for }  - \frac{3\sqrt{5}}{7} \leq \eta \leq \frac{3\sqrt{5}}{7},\\
        U_S(0)=U_S(1)=\sigma_1^x &\text{for } \frac{3\sqrt{5}}{7}\leq \eta < 1.
    \end{cases}
\end{align}
Note that when $-1<\eta<-\frac{3\sqrt{5}}{7}$ and $\frac{3\sqrt{5}}{7}<\eta<1$, the operations on qubit 1 that maximize $E_\mathrm{ext}$ is not feedback control. In this region, $\rho_S^\mathrm{i}$ is almost equal to $\ket{\downarrow}_1 \bra{\downarrow}_1$ or $\ket{\uparrow}_1 \bra{\uparrow}_1$, and thus the optimal operations for extracting energy is to do nothing or the spin flip.
Next, we calculate the upper bound presented in Eq.~(13) in the main text. The entanglement entropy of the qubit 1 is
\begin{align}
    S(\rho_S^\mathrm{i})&=\mathrm{tr}\rho_S^\mathrm{i} \log \rho_S^\mathrm{i}\nonumber\\&=-\frac{1+\eta}{2}\log \frac{1+\eta}{2}-\frac{1-\eta}{2}\log \frac{1-\eta}{2}.
\end{align}
After the measurement $P_A(\mu)$, the entanglement entropy of qubit 1 can decrease on average to
\begin{align}
    \sum_\mu p_\mu S(\rho_S^\mathrm{m}(\mu))=&-\sum_\mu p_\mu \mathrm{tr} \rho_S^\mathrm{m}(\mu) \log \rho_S^\mathrm{m}(\mu)\nonumber\\
    =&-\sum_\mu p_\mu [\lambda_\uparrow(\mu) \log \lambda_\uparrow(\mu) + \lambda_\downarrow(\mu) \log \lambda_\downarrow(\mu)]\nonumber\\
    =&-\frac{1+\eta}{2} \log \frac{1+\eta}{2}  -\frac{1-\eta}{2} \log \frac{1-\eta}{2}\nonumber\\ &-\frac{1}{14}\left(7+3(2n_x+n_z)\right) \log \left(7+3(2n_x+n_z)\right) -\frac{1}{14}\left(7-3(2n_x+n_z)\right) \log \left(7-3(2n_x+n_z)\right) \nonumber\\
    &+ \frac{1}{14}\left(7+3\eta(2n_x+n_z)\right) \log \left(7+3\eta(2n_x+n_z)\right)  +\frac{1}{14}\left(7-3\eta(2n_x+n_z)\right) \log \left(7-3\eta(2n_x+n_z)\right)
    \nonumber\\
    \geq& -\frac{1+\eta}{2} \log \frac{1+\eta}{2}  -\frac{1-\eta}{2} \log \frac{1-\eta}{2}\nonumber\\ &-\frac{1}{14}\left(7+3\sqrt{5}\right) \log \left(7+3\sqrt{5}\right) -\frac{1}{14}\left(7-3\sqrt{5}\right) \log \left(7-3\sqrt{5}\right) \nonumber\\
    &+ \frac{1}{14}\left(7+3\sqrt{5}\eta\right) \log \left(7+3\sqrt{5}\eta\right)  +\frac{1}{14}\left(7-3\sqrt{5}\eta\right) \log \left(7-3 \sqrt{5}\eta\right)\nonumber\\
    =& E_\mathrm{F}^{S\textit{-}E}(\rho_{SE}^\mathrm{i}),\label{eq:aveEE}
\end{align}
where we used $-\sqrt{5}\le 2n_x+n_z \le \sqrt{5}$. From Eq.~(S.\ref{eq:aveEE}), QC-mutual information $I_\mathrm{QC}\equiv S(\rho_S^\mathrm{i})-\sum_\mu p_\mu S(\rho_A^\mathrm{m}(\mu))$ and the entanglement $\overleftarrow{\mathcal{E}}_{SA}\equiv S(\rho_S^\mathrm{i})-E_\mathrm{F}^{S\textit{-}E}(\rho_{SE}^\mathrm{i})$ can be computed as
\begin{align}
    I_\mathrm{QC}=&\frac{1}{14}\left(7+3(2n_x+n_z)\right) \log \left(7+3(2n_x+n_z)\right) \frac{1}{14}\left(7-3(2n_x+n_z)\right) \log \left(7-3(2n_x+n_z)\right) \nonumber\\
    &- \frac{1}{14}\left(7+3\eta(2n_x+n_z)\right) \log \left(7+3\eta(2n_x+n_z)\right)  -\frac{1}{14}\left(7-3\eta(2n_x+n_z)\right) \log \left(7-3\eta(2n_x+n_z)\right),\\
    \overleftarrow{\mathcal{E}}_{SA}=&\frac{1}{14}\left(7+3\sqrt{5}\right) \log \left(7+3\sqrt{5}\right) +\frac{1}{14}\left(7-3\sqrt{5}\right) \log \left(7-3\sqrt{5}\right) \nonumber\\
    &- \frac{1}{14}\left(7+3\sqrt{5}\eta\right) \log \left(7+3\sqrt{5}\eta\right)  -\frac{1}{14}\left(7-3\sqrt{5}\eta\right) \log \left(7-3 \sqrt{5}\eta\right).
\end{align}
In our effective thermodynamics, the thermal equilibrium state $\sigma_S$ is defined by Eq.~(3) in the main text. Thus, the von Neumann entropy of $\sigma_S$ is
\begin{align}
    S(\sigma_S)&=-\mathrm{tr}\sigma_S \log \sigma_S\nonumber\\
    &=\frac{\log {(1+e^{2\beta_\mathrm{eff}})}}{1+e^{2\beta_\mathrm{eff}}}+\frac{\log {(1+e^{-2\beta_\mathrm{eff}})}}{1+e^{-2\beta_\mathrm{eff}}},
\end{align}
and the KL divergence $D(\rho_S^\mathrm{i}||\sigma_S)$ is calculated as
\begin{align}
    D(\rho_S^\mathrm{i}||\sigma_S)&=\mathrm{tr}\rho_S^\mathrm{i} \log \sigma_S -\mathrm{tr} \rho_S^\mathrm{i} \log \sigma_S\nonumber\\
    &=-\log 2 + \frac{1+\eta}{2} \log (1+\eta)(1+e^{2\beta_\mathrm{eff}})+ \frac{1-\eta}{2}\log (1-\eta)(1+e^{-2\beta_\mathrm{eff}}).
\end{align}
The effective inverse temperature $\beta_\mathrm{eff}$ can be determined numerically by Eq.~(12) in the main text. When $\eta=0$,  $\beta_\mathrm{eff}$ can be calculated analytically as $\beta_\mathrm{eff}=\frac{1}{2}\log{\frac{7+3\sqrt{5}}{7-3\sqrt{5}}}$. At this value of $\eta$, $D(\rho_S^\mathrm{i}||\sigma_S)=-\log 2 +\log 7$, $\overleftarrow{\mathcal{E}}_{SA}=\log 2 -\log 7 + \frac{3\sqrt{5}}{14}\log \frac{7+3\sqrt{5}}{7-3\sqrt{5}}$, and thus $\frac{1}{\beta_\mathrm{eff}}\left[D(\rho_S^\mathrm{i}||\sigma_S)+\overleftarrow{\mathcal{E}}_{SA}\right]=\mathrm{max}E_\mathrm{ext}=\frac{3\sqrt{5}}{7}$. Therefore, in this model, $E_\mathrm{ext}$ reaches the upper bound if the eigenvalues of $\rho_S^\mathrm{m}(\mu)$ are independent of $\mu$.

\end{widetext}

\end{document}